\documentclass[12pt,preprint]{aastex}

\shorttitle{Polarization of Sgr A*}
\shortauthors{Huang et al.}

\begin{document}


\title{Polarized Emission of Sagittarius A*}



\author{Lei Huang\altaffilmark{1,3,4}, Siming Liu\altaffilmark{2},
Zhi-Qiang Shen\altaffilmark{3}, Ye-Fei Yuan\altaffilmark{1}, Mike
J. Cai\altaffilmark{4}, Hui Li\altaffilmark{5}, and Christopher
L. Fryer\altaffilmark{5, 6} } 

\altaffiltext{1}{Key Laboratory for Research in Galaxies and
Cosmology, The University of Sciences and Technology of China,
Chinese Academy of Sciences, Hefei 230026, China;
mlhuang@ustc.edu.cn, yfyuan@ustc.edu.cn}

\altaffiltext{2}{Department of Physics and Astronomy, University
of Glasgow, Glasgow G12 8QQ, UK; sliu@astro.gla.ac.uk}

\altaffiltext{3}{Key Laboratory for Research in Galaxies and
Cosmology, Shanghai Astronomical Observatory, Chinese Academy of
Sciences, Shanghai 200030, China; zshen@shao.ac.cn}

\altaffiltext{4}{Academia Sinica, Institute of Astronomy and
Astrophysics, Taipei 106, Taiwan; mike@asiaa.sinica.edu.tw}

\altaffiltext{5}{Los Alamos National Laboratory, Los Alamos, NM
87545; hli@lanl.gov, fryer@lanl.gov}

\altaffiltext{6}{Physics Department, The University of Arizona,
Tucson, AZ 85721}


\begin{abstract}

We explore the parameter space of the two temperature pseudo-Newtonian
 Keplerian accretion flow model for the millimeter and shorter
 wavelength emission from Sagittarius A*. A general relativistic
 ray-tracing code is used to treat the radiative transfer of polarized
 synchrotron emission from the flow. The synchrotron self-Comptonization
 and bremsstrahlung emission components are also included. It is shown that
 the model can readily account for the millimeter to sub-millimeter
 emission characteristics with an accretion rate of $\sim 6 \times
 10^{17} \mathrm{g} \cdot \mathrm{s}^{-1}$ and an inclination angle of 
 $\sim 40^\circ$. However, the corresponding model predicted 
 near-infrared and X-ray fluxes are more than one order of magnitude
 lower than the observed `quiescent' state values. While the
 extended quiescent-state X-ray emission has been attributed to thermal
 emission from the large-scale accretion flow, the NIR emission and
 flares are likely dominated by emission regions either within the last
 stable orbit of a Schwarzschild black hole or associated 
 with outflows. With the viscous parameter derived from numerical
 simulations, there is still a degeneracy between the electron heating
 rate and the magnetic parameter.  A fully general
 relativistic treatment with the black hole spin incorporated will
 resolve these issues.

\end{abstract}


\section{INTRODUCTION}
\label{intro}

Sagittarius (Sgr) A*, the compact radio source associated with the
super-massive black hole at the Galactic center, is perhaps the
best source for the study of physical processes near the event
horizon of a black hole \citep{Scho02,Ghez05}. 
Although its luminosity is relatively low, with the high
resolution and sensitivity of modern instruments, the source has
been routinely observed from radio to X-rays. It may also play a
role in the production of TeV gamma-ray emission from the Galactic
center region \citep{Liu06,Ahar06}. 
The low luminosity also renders the source optically thin at millimeter
and shorter wavelengths. One therefore can observe emission from
the very inner region close to the black hole directly at these
wavelengths.

Most of Sgr A* emission is released in the sub-millimeter band,
where a strong linear polarization is observed. The flux density
also varies with the variation amplitude and rate decreasing with
the decrease of the observation frequency \citep{Herr04}. 
The variation timescale of sub-millimeter emission can be
as short as a few hours, slightly longer than the dynamical time
near the black hole. The source is highly variable at shorter
wavelengths with the peak luminosity of some near infrared (NIR)
and X-ray flares comparable to the sub-millimeter luminosity
\citep{Baga01,Genz03} 
The variation
timescale of these flares is consistent with events occurring
within a few Schwarzschild radii of the black hole. And correlated
flare activities in the X-ray, NIR, millimeter, and radio bands
suggest outflows during some flares
\citep{Zhao04,Yuse07,Marr07,Ecka08}. 
The flares also contain rich temporal, spectral, and polarization information 
\citep{Porq08,Fala08,Ecka08}. 
Detailed modelling of these emission characteristics is expected
to probe the geometry near the black hole, processes of the
general relativistic magnetohydrodynamics (GRMHD) \citep{GMT03}, 
and the kinetics of electron heating and acceleration in a
magnetized relativistic plasma \citep{LPM04, lfl07}. 

Motivated by MHD simulations of radiatively inefficient accretion
flows \citep{HB91,HB02}, we proposed a
Keplerian accretion flow model for the time averaged millimeter
and shorter wavelength emission from Sgr A* \citep{MLC00,MLC01}. 
The model plays important roles in constraining the time
averaged properties of the plasma near the black hole \citep{YQN03} 
for detailed GRMHD studies \citep{Nobl07}. 
It also treats the polarization characteristics quantitatively \citep{BML01}. 
The model is later generalized to a two temperature
accretion flow in a pseudo-Newtonian potential \citep{Liu07, lfl07}. 
\citet{Huang08} first carried out a self-consistent treatment
of the transfer of polarized emission through the accretion flow
and predicted distinct linear and circular polarization
characteristics between the sub-millimeter and NIR band due to
general relativistic light bending and birefringence effects. In
this paper, we present the details of the calculations (\S\
\ref{radia}), explore the model parameter space (\S\
\ref{results}), and discuss future developments (\S\
\ref{discuss}).

\section{RADIATIVE TRANSFER OF SYNCHROTRON RADIATION}
\label{radia}

Although the importance of a self-consistent treatment of
polarized radiative transfer through relativistic plasmas near
black holes was recognized a while ago \citep{Melr97}, 
there are still significant uncertainties in the quantitative details
\citep{Shch08}. The thermal synchrotron emission and
absorption coefficients have been derived independently by several
authors \citep{LW68,Sazo69,Melr71}, 
which in general show agreement. The derivation of the Faraday rotation
and conversion coefficients was mostly done by \citep{Melr97}. 
\citet{Shch08} recently derived expressions appropriate for
the transition from non-relativistic temperatures to relativistic
ones. In this section, we have a brief discussion of the general
theory of polarized radiation transfer and show that the Faraday
rotation and conversion coefficients are not independent. The
radio of the two is equal to the ratio of the circular and linear
emission coefficients. Therefore our approach of deriving the
Faraday conversion coefficient from the emission coefficients and
Faraday rotation coefficient is appropriate \citep{Huang08}. 

\subsection{Properties and Transfer of Partially Polarized Radiation}

\begin{figure}[ht]
\vspace{-0mm}
\begin{center}
\includegraphics[width=14cm]{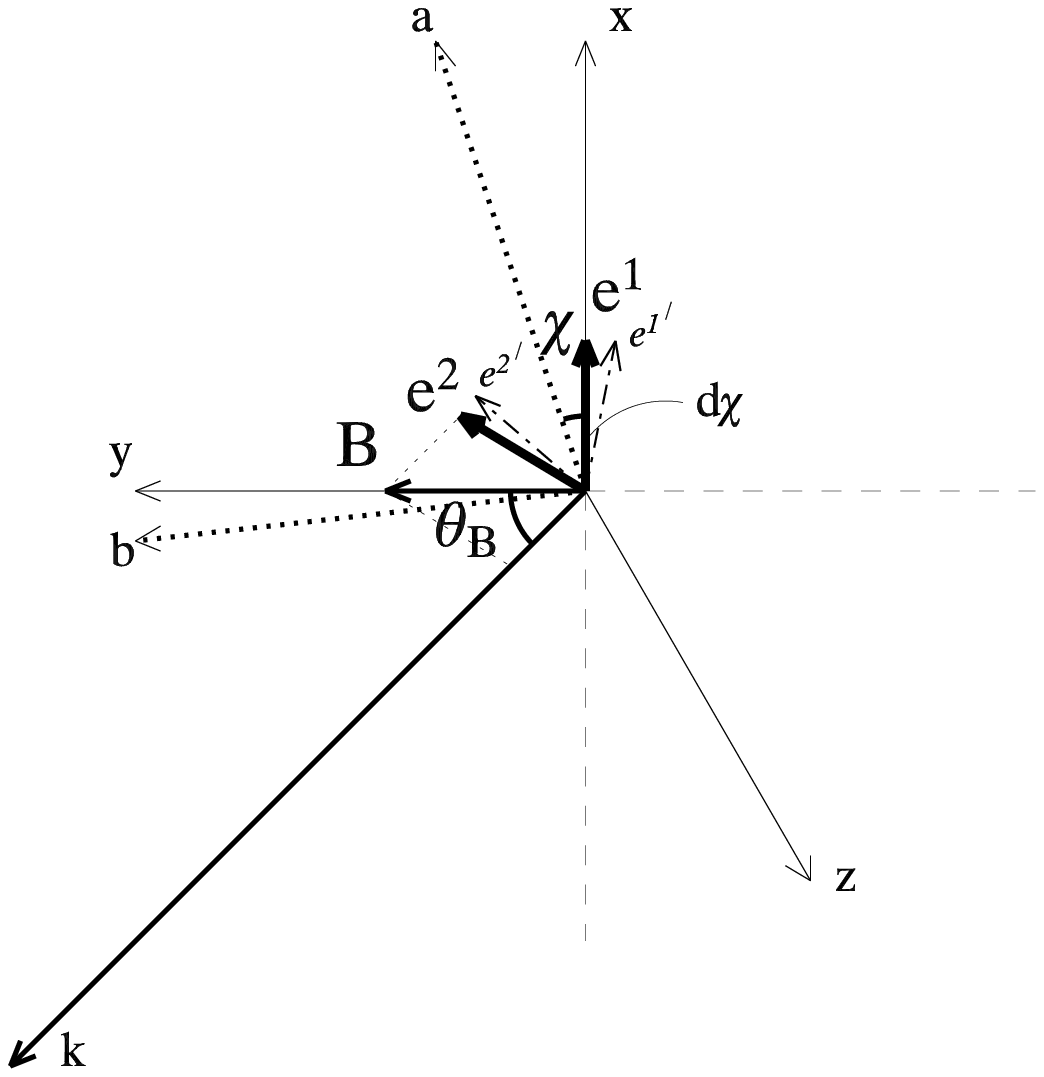}
\vspace{-5mm}\caption{ Sketch map of polarization in synchrotron
radiation: vectors ${\rm k}$, ${\rm B}$, ${\rm e}^1$, and ${\rm
e}^2$ represent the wave vector, magnetic field, e-mode, and
o-mode, respectively. $\chi$ and $d\chi$ respectively represent
the EVPA defined in arbitrary coordinate system $(a,b)$ and the
change of EVPA due to radiation transfer effect. See text for detail.
\label{coord}}
\end{center}
\end{figure}

The synchrotron radiation from an individual particle is
elliptically polarized. The major axis of the polarization ellipse
{\bf e}$^1$ is perpendicular to the plane spanned by the magnetic
vector {\bf B} and the wave vector {\bf k}, and the minor axis
{\bf e}$^2\propto$ {\bf k}$\times${\bf e}$^1$ is perpendicular to
the major axis and wave vector. The electric field component of
the radiation can be projected along the major and minor axes,
namely the e- and o-component, respectively. Figure \ref{coord}
shows the corresponding geometry in the three dimensional (3D)
Cartesian coordinates $(x,y,z)$:
\begin{eqnarray}
  {\rm \bf B}/B &=& (0,\,1,\,0), \qquad
{\rm \bf k}/k \quad=\quad (0,\,\cos\theta_B,\,\sin\theta_B),  \nonumber \\
  {\rm \bf e}^1 &=& (1,\,0,\,0), \qquad\quad
{\rm \bf e}^2 \quad=\quad (0,\,\sin\theta_B,\,-\cos\theta_B),
\end{eqnarray}
where $\theta_B$ is the angle between {\bf B} and {\bf k}, and $k$
and $B$ are the wave number and the amplitude of the magnetic
field, respectively. For a population of relativistic particles
with a smooth pitch angle distribution, the circular polarization
(CP) component of synchrotron emission almost cancels out, and the
emissivity of the e-component is generally much greater than that
of the o-component. Therefore, the synchrotron radiation from a
population of relativistic particles is often treated as partially
linearly polarized along {\bf e}$^1$. The e-component base ${\rm
\bf e}^1$ then indicates the electric vector of the linearly
polarized (LP) radiation. Since the CP component does not cancel
out exactly, sometimes it is more convenient to use the
right-handed and left-handed CP bases
\begin{eqnarray} \label{CPmodes}
  \mathrm{\bf e}^R &=& \frac{1}{\sqrt{2}}(\mathrm{\bf e}^1 +
   i\mathrm{\bf e}^2),
   \qquad \mathrm{\bf e}^L \quad=\quad \frac{1}{\sqrt{2}}(\mathrm{\bf e}^1 -
   i\mathrm{\bf e}^2).
\end{eqnarray}

According to the intensity matrix $\mathcal{I}^{ij}$ defined by
\citet{Melr71} with an arbitrary orthogonal bases, the total
intensity is given by
\begin{eqnarray}
  I &=& \mathrm{trace}(\mathcal{I}^{ij}) \quad=\quad
   \mathcal{I}^{11}+\mathcal{I}^{22} \quad=\quad
   \mathcal{I}^{RR}+\mathcal{I}^{LL}.
\end{eqnarray}
Then one has the formal polarization vector
$\vec{p}=(p_Q,p_U,p_V)$ introduced by \citet{LL04}
\begin{eqnarray}
  p_Q &=&
   \frac{\mathcal{I}^{11}-\mathcal{I}^{22}}{\mathcal{I}^{11}+\mathcal{I}^{22}},
   \nonumber \\
  p_U &=&
   \frac{\mathcal{I}^{12}+\mathcal{I}^{21}}{\mathcal{I}^{11}+\mathcal{I}^{22}},
   \nonumber \\
  p_V &=&
  \frac{i(\mathcal{I}^{12}-\mathcal{I}^{21})}{\mathcal{I}^{11}+\mathcal{I}^{22}}
   \quad=\quad
   \frac{\mathcal{I}^{RR}-\mathcal{I}^{LL}}{\mathcal{I}^{RR}+\mathcal{I}^{LL}},
\end{eqnarray}
The total polarization fraction
$\Pi$, CP fraction $\Pi_C$, LP fraction $\Pi_L$, and the electric vector
position angle (EVPA) $\chi_0$ are then given by
\begin{eqnarray}
  \Pi &=& \left[ 1 - \frac{4
       \mathrm{det}(\mathcal{I}^{ij})}{\mathrm{trace}(\mathcal{I}^{ij})^2}
      \right]^{1/2},   \nonumber \\
  \Pi_C &=& p_V,   \nonumber \\
  \Pi_L &=& (\Pi^2 - \Pi_C^2)^{1/2} \quad=\quad (p_Q^2 + p_U^2)^{1/2},
   \nonumber \\
\tan 2\chi_0 &=& \frac{p_U}{p_Q}\,.
\end{eqnarray}
The normalized polarization vector can be rewritten in the form of
the 4D Stokes vector as $\vec{S}=(I,Q,U,V)^T=I(1, p_Q, p_U,
p_V)^T$. The Stokes vector, the intensity matrix, and ($I$,
$\Pi_C$, $\Pi_L$, $\chi_0$) all give a complete description of the
properties of partially polarized emission.

In order to describe the polarized radiative transfer in
highly-magnetized plasma, \citet{LL04} introduce another three 3D
formal vector $\vec{\epsilon}, \vec{\eta}, \vec{\rho}$, where
$\vec{\epsilon}$ is the normalized emission vector, which is
related to the 4D emission coefficient
$\vec{\varepsilon}=(\varepsilon_I, \varepsilon_Q, \varepsilon_U,
\varepsilon_V)^T$:  $\vec{\epsilon} \equiv (\epsilon_Q,
\epsilon_U, \epsilon_V) = (\varepsilon_Q, \varepsilon_U,
\varepsilon_V) /I$, $\vec{\eta} \equiv (\eta_Q, \eta_U, \eta_V)$
is the absorption vector, and $\vec{\rho}$ is the Faraday rotation
vector. The total emission and average absorption coefficients are
represented by $\varepsilon_I=\epsilon_I I$ and $\eta_I$,
respectively. Then the transfer of polarized emission is described
with
\begin{eqnarray}
\label{transfer}
  {{\rm d} I\over {\rm d} \rm s} &=& - (\eta_I + \vec{\eta} \cdot
  \vec{p}-\epsilon_I)I, \nonumber \\
  {{\rm d} \vec{p}\over {\rm d} \rm s} &=& -\vec{\eta} + (\vec{\eta}
   \cdot{}
   \vec{p})\vec{p}+\vec{\rho}\times\vec{p}+\vec{\epsilon}-\epsilon_I\vec{p}\,,
\end{eqnarray}
which govern the evolution of the polarization properties of
radiation through a magnetized plasma. These equations can be
rewritten in the form of Stokes vector as
\begin{eqnarray}
\label{transferS}
  \frac{\rm d}{\rm d \rm s}\vec{S} &=& \vec{\varepsilon} \quad-\quad
  \mathbf{K} \vec{S},
\end{eqnarray}
or explicitly
\begin{eqnarray}
  \frac{\rm d}{\rm d \rm s} \left( \begin{array}{cccc}
    I \\ Q \\ U \\ V
  \end{array} \right) &=&
  \left( \begin{array}{cccc}
    \varepsilon_I \\ \varepsilon_Q \\ \varepsilon_U \\ \varepsilon_V
  \end{array} \right) \quad-\quad
  \left( \begin{array}{cccc}
    \eta_I & \eta_Q & \eta_U & \eta_V \\
    \eta_Q & \eta_I & \rho_V & -\rho_U \\
    \eta_U & -\rho_V & \eta_I & \rho_Q \\
    \eta_V & \rho_U & -\rho_Q & \eta_I
  \end{array} \right)
  \left( \begin{array}{cccc}
    I \\ Q \\ U \\ V
  \end{array} \right).
\end{eqnarray}
One can use the rotation matrix
\begin{eqnarray}
  \mathrm{R}(\chi) &=&
  \left( \begin{array}{cccc}
    1 & 0 & 0 & 0 \\
    0 & \cos 2\chi & \sin 2\chi & 0 \\
    0 & -\sin 2\chi & \cos 2\chi & 0 \\
    0 & 0 & 0 & 1
  \end{array} \right),
\label{RMatrix}
\end{eqnarray}
to transform the Stokes vector from the coordinates $({\rm \bf e}^1,{\rm
\bf e}^2)$ to the reference coordinates $({\bf a},{\bf b})$, where {\bf a} corresponds
to the North at the observer and {\bf b} corresponds to the East (see Figure
\ref{coord}). Then we have
\begin{eqnarray}
  \frac{\rm d}{\rm d \rm s}\vec{S}^\prime &=& \vec{\varepsilon}^\prime
   \quad-\quad \mathbf{K}^\prime \quad \vec{S}^\prime\,,
\end{eqnarray}
where $\vec{S}^\prime = \mathrm{R}(\chi)\vec{S}$,
$\vec{\varepsilon}^\prime= \mathrm{R}(\chi)\vec{\varepsilon}$, and
$\mathbf{K}^\prime = \mathrm{R}(\chi) \mathbf{K} \mathrm{R}(-\chi)$.

\subsection{Synchrotron Emission, Absorption, and Faraday Coefficients}

For a population of relativistic electrons in the Maxwell distribution,
i.e., relativistic thermal distribution, with the number density $N_0$ and
temperature $T_e\gg m_ec^2/k$, the distribution function with respect to the
electron energy $E$ is
\begin{eqnarray}
  N(E) &=& N_0 \frac{E^2}{2(kT_e)^3} \mathrm{exp}(-E/kT_e)\,,
\end{eqnarray}
where $m_e$, $c$, and $k$ are the electron mass, speed of light, and
Boltzmann constant, respectively.
In the coordinates of ({\bf e}$^1$, {\bf e}$^2$), the four synchrotron
emission coefficients are \citep{Sazo69,Pach70,Melr71}
\begin{eqnarray}
\label{emission}
  \varepsilon_I &=& \frac{e^2 m_e^2 c^3}{\sqrt{3}} \frac{N_0}{2
   (kT_e)^2} \nu I_I(x_M),  \nonumber \\
  \varepsilon_Q &=& \frac{e^2 m_e^2 c^3}{\sqrt{3}} \frac{N_0}{2
   (kT_e)^2} \nu I_Q(x_M),  \nonumber \\
  \varepsilon_U &=& 0,  \nonumber \\
  \varepsilon_V &=& \frac{4 e^2 m_e^3 c^5 \mathrm{cot}\theta_B}{3
   \sqrt{3}} \frac{N_0}{2 (kT_e)^3} \nu I_V(x_M)\,,
\end{eqnarray}
where $e$ is the elementary charge units and $x_M$ is the ratio of the
emission frequency $\nu$ to the characteristic frequency $\nu_c={3eB
\mathrm{sin}\theta_B (kT_e)^2}/{4\pi m_e^3 c^5}$,  and the
integrations $I_I, I_Q$ and $I_V$ can be calculated with the aid of the
modified Bessel functions $K_\alpha(z)$:
\begin{eqnarray}
  F(x) &=& x \int_x^{\infty} K_{5/3}(z)dz\,,  \nonumber \\
  G(x) &=& x K_{2/3}(x)\,,  \nonumber \\
  H(x) &=& x K_{1/3}(x) + \int_x^{\infty} K_{1/3}(z)dz, \nonumber
   \quad \nonumber \\
  I_I(x_M) &=& \frac{1}{x_M} \int_0^{\infty} z^2 \mathrm{exp}(-z)
   F\left(\frac{x_M}{z^2}\right)  dz\,,   \nonumber \\
  I_Q(x_M) &=& \frac{1}{x_M} \int_0^{\infty} z^2 \mathrm{exp}(-z)
   G\left(\frac{x_M}{z^2}\right)  dz\,,   \nonumber \\
  I_V(x_M) &=& \frac{1}{x_M} \int_0^{\infty} z^{\quad} \mathrm{exp}(-z)
   H\left(\frac{x_M}{z^2}\right)  dz\,.
\end{eqnarray}
The emission coefficients along the two axes ${\rm \bf e}^1$ and ${\rm
\bf e}^2$ are given, respectively, by
\begin{eqnarray}
  \varepsilon_1 &=& \frac{\varepsilon_I + \varepsilon_Q}{2},  \nonumber \\
  \varepsilon_2 &=& \frac{\varepsilon_I - \varepsilon_Q}{2}.
\end{eqnarray}
The absorption coefficient can be calculated with the Kirchhoff's law
\begin{eqnarray}
\label{source}
\frac{S}{2} &=&
\frac{\varepsilon_1}{\eta_1}=\frac{\varepsilon_2}{\eta_2}\,, \nonumber
\\
\eta_I &=& \frac{\eta_1+\eta_2}{2}\,,\nonumber \\
\vec{\eta} &=&   \frac{(\varepsilon_Q,\,
 \varepsilon_U,\,\varepsilon_V)}{S}=\frac{\vec{\epsilon}I}{S}\,,
\end{eqnarray}
where $S=\varepsilon_I/\eta_I$ is the source function,
$S=B_\nu=2h\nu^3/c^2[\exp{(h\nu/kT_e)}-1]$ for a thermal distribution,
and $h$ is the Planck constant.

According to \citet{Melr97}, the high-frequency waves may be treated as
two transverse nature wave modes with dispersion relations
$$k_\pm^2c^2-\omega^2 = \frac{1}{2} \{ \alpha^{11} + \alpha^{22} \pm
[(\alpha^{11} - \alpha^{22})^2 + 4\alpha^{12}\alpha^{21}]^{1/2}
\},$$ where $\omega=2\pi\nu$, $k$ is the wave number, and the
response tensor $\alpha^{ij}$ satisfies
$\alpha^{12}=-\alpha^{21}$. The polarization vectors of the two
natural modes are then given by
\begin{eqnarray}
    {\rm \bf e}^{\pm} = \frac{T_\pm {\rm \bf e}^1+ i {\rm \bf
     e}^2}{\sqrt{1 + T_\pm^2}}, &\quad& T_\pm = \frac{\alpha^{11} -
     \alpha^{22} \mp  [(\alpha^{11} - \alpha^{22})^2 +
     4\alpha^{12}\alpha^{21}]^{1/2}}{2i \alpha^{12}},
\end{eqnarray}
which are two orthogonal elliptically polarized modes with axial ratios
$T_\pm$.  Note that $T_+T_- = -1$ implying that the two natural modes
have identical ellipticity and opposite handedness. The electric vector
${\rm \bf E}$ of any
wave can be decomposed on the bases of the two natural modes as $E_+
{\rm \bf e}^+ + E_- {\rm \bf e}^-$ or on the bases of the two LP modes
as $E_1 {\rm \bf e}^1 + E_2 {\rm \bf e}^2$. Then
$$E_1 = E_+ \frac{T_+}{\sqrt{1 + T_+^2}} + E_- \frac{T_-}{\sqrt{1 + T_-^2}}$$
and
$$E_2 = E_+ \frac{1}{\sqrt{1 + T_+^2}} + E_- \frac{1}{\sqrt{1 + T_-^2}}.$$
Using the definitions in equation (\ref{CPmodes}), a third decomposition can be
made on two CP modes as $E_R {\rm e}^R + E_L {\rm e}^L$, with
$$E_R = E_+ \frac{T_+ + 1}{\sqrt{2(1 + T_+^2)}} + E_- \frac{T_- +
1}{\sqrt{2(1 + T_-^2)}}$$
and
$$E_L = E_+ \frac{T_+ - 1}{\sqrt{2(1 + T_+^2)}} + E_- \frac{T_- -
1}{\sqrt{2(1 + T_-^2)}}.$$
The total emission coefficient $\varepsilon_I$ is proportional to $E^2$, i.e.,
\begin{eqnarray}
    \varepsilon_I &=& \Bigg\{ \begin{array}{ccc}
        \varepsilon_+ + \varepsilon_- \quad\propto\quad E_+^2 +
         E_-^2 \\
        \varepsilon_1 + \varepsilon_2 \quad\propto\quad E_1^2 +
         E_2^2 \\
        \varepsilon_R + \varepsilon_L \quad\propto\quad E_R^2 + E_L^2
    \end{array}.
\end{eqnarray}
Then the LP and CP emission coefficients ($\varepsilon_Q$ and $\varepsilon_V$)
are given by
\begin{eqnarray}
    \varepsilon_Q &=& \varepsilon_1 - \varepsilon_2 =
     \varepsilon_{Q+} + \varepsilon_{Q-} = \frac{T_+^2 -1}{1+T_+^2}
     \varepsilon_+ +  \frac{T_-^2 -1}{1+T_-^2}\varepsilon_- =
     \frac{T_++T_-}{T_+-T_-}(\varepsilon_+-\varepsilon_-)
     \nonumber \\
    \varepsilon_V &=& \varepsilon_R - \varepsilon_L =
     \varepsilon_{V+} + \varepsilon_{V-}= \frac{2
     T_+}{1+T_+^2}\varepsilon_+ +  \frac{2
     T_-}{1+T_-^2}\varepsilon_-
     =\frac{2}{T_+-T_-}(\varepsilon_+-\varepsilon_-),
\end{eqnarray}
which implies that
\begin{eqnarray}
\label{parallel}
    \frac{\varepsilon_V}{\varepsilon_Q} &=& \frac{2}{T_++T_-}=
     \frac{2i\alpha^{12}}{\alpha^{11} - \alpha^{22}} =
     \frac{\rho_{\rm RM}}{\rho_{\rm RRM}},
\end{eqnarray}
where $\rho_{\rm RM}=i\alpha^{12}/2\omega c$ is called the rotation
measure, or the Faraday rotation coefficient, and $\rho_{\rm
RRM}=(\alpha^{11}-\alpha^{22})/4\omega c$ is called the relativistic
rotation measure, or the Faraday conversion coefficient. Note that
$\varepsilon_\pm^2=\varepsilon_{Q\pm}^2+\varepsilon_{V\pm}^2$,
$\varepsilon_\pm/\eta_\pm = S/2$,
$\eta_{Q}=\varepsilon_{Q}/S = (\eta_{Q+}+\eta_{Q-})/2$,
$\eta_{V}=\varepsilon_{V}/S = (\eta_{V+}+\eta_{V-})/2$.

In a uniform plasma,  $\vec{\rho} = (2\rho_{\rm RRM}, 0, 2\rho_{\rm
RM})$ and the three formal vector $\vec{\epsilon}$, $\vec{\eta}$, and
$\vec{\rho}$ are parallel to each other. If there is no external
radiation,  the term  $(\vec{\rho} \times \vec{p})$ in equations
(\ref{transfer}) vanishes \citep{LL04,Beke66,KM98} and one has the
solution with $\vec{p}$ also parallel to these vectors. Then equations
(\ref{transfer}) become
\begin{eqnarray}
\label{transfer0}
  {{\rm d} I\over {\rm d} \rm s} &=& - [(1+\Pi\Pi_0)\eta_I -\epsilon_I]I
  =-\left[(1+\Pi\Pi_0)\frac{I}{S}-1\right] \varepsilon_I\,,
  \nonumber \\
  {{\rm d} \Pi\over {\rm d} \rm s} &=&
   \eta_I\Pi_0(\Pi^2-1)+\epsilon_I(\Pi_0-\Pi)=\left[\frac{\Pi_0(\Pi^2-1)}{S}
                + \frac{\Pi_0-\Pi}{I}\right]\varepsilon_I\,,
\end{eqnarray}
where
$\Pi_0=(\varepsilon_Q^2+\varepsilon_U^2+\varepsilon_V^2)^{1/2}/\varepsilon_I$
is the polarization fraction of an optically thin source. Note that in
this case, if we denote the unit vector along $\vec{\epsilon}$ as
$\vec{u}$, then $\vec{p}=\Pi\vec{u}$,
$\vec{\epsilon}= \Pi_0\epsilon_I \vec{u}$,
$\vec{\eta}=\vec{\epsilon}I/S=\Pi_0\eta_I\vec{u}$.  With the
increase of the optical depth, we see that $\Pi$ decreases, and the
existence of $\Pi$ enhances the self-absorption by a factor of $1+\Pi\Pi_0$.

Therefore, in a uniform plasma, besides the source function that is
determined by the particle distribution, there are four independent
coefficients. From the three emission coefficients given by equations
(\ref{emission}) and the Faraday rotation coefficient $\rho_V=2\rho_{\rm
RM}$, one can derive the absorption and Faraday conversion coefficients
with equations (\ref{source}) and (\ref{parallel}),
respectively. \citet{Melr97} derived the three emission
coefficients and two Faraday coefficients separately. It appears that
his results are not exactly in agreement with equation
(\ref{parallel}). Neither is the  $\rho_{\rm RM}$ and
$\rho_{\rm RRM}$ obtained by \citet{Shch08}. All these calculations
invoke some approximations to simplify the Trubnikov's linear response
tensor. While equation (\ref{parallel}) is derived from very general
theoretical considerations without any approximations. We therefore stand
by our approach of deriving $\rho_{\rm RRM}$ from $\rho_{\rm RM}$ and
the emission coefficients in a previous paper \citep{Huang08}.

\citet{Melr97} gives $\rho_{\rm RM}$ for thermal plasmas in the
cold ($\gamma_C = kT_e/m_ec^2+1\approx 1$) and extremely
relativistic (ER) ($\gamma_C \gg 1$) limits:
\begin{eqnarray}
  \rho_{\rm RM}^{{\rm C}} &=& \frac{e^3 N_0 B \cos\theta_B}{2 \pi
   \gamma_c m_e^2 c^2 \nu^2},   \nonumber \\
  \rho_{\rm RM}^{{\rm ER}} &=& \frac{\ln\gamma_c}{2\gamma_c} \frac{e^3
   N_0 B \cos\theta_B}{2 \pi \gamma_c m_e^2 c^2 \nu^2}.
\end{eqnarray}
We use the following extrapolation to obtain $\rho_{\rm RM}$ for
arbitrary electron temperatures
\begin{eqnarray}
  \rho_{\rm RM} &=& \gamma_c^{-1} (\rho_{\rm RM}^{{\rm C}} - \rho_{\rm
   RM}^{{\rm ER}}) + \rho_{\rm RM}^{{\rm ER}}\,,
\end{eqnarray}
which is simple and accurate enough according to \citet{Shch08}.
For cold plasmas, cyclotron emission dominates, $\varepsilon_I >
\varepsilon_V \gg \varepsilon_Q$, and $\rho_V \gg \rho_Q$. For hot
plasmas, synchrotron emission dominates, $\epsilon_I > \epsilon_Q
\gg \epsilon_V$, and $\rho_Q \gg \rho_V$. 

For a given ray, the magnetic field structure of the accretion
flow determines the coordinates ({\bf e}$^1$, {\bf e}$^2$). For
given reference directions of the North and East, we also obtain
the coordinates ({\bf a}, {\bf b}). One therefore can use the
coefficients obtained above in the coordinates of ({\bf e}$^1$,
{\bf e}$^2$) to obtain the contributions of this ray to the Stokes
parameters in the coordinates of ({\bf a}, {\bf b}).

\subsection{General Relativistic Radiative Transfer}

In magnetized plasmas around black holes, photons experience both
the general relativistic light bending and plasma birefringence
effects and change their propagation directions.  \citet{Fant97}
showed how photons propagate along the null geodesics near black
holes. \citet{BB03} discussed how the plasma effect causes
refraction, which makes the photon propagation path deviate from
the null geodesics. However, such refraction is significant only
if either the electron cyclotron frequency $\nu_B$ or the electron
plasma frequency $\nu_P$ is comparable to the observation
frequency $\nu_\mathrm{obs}$. In the observational band (from
millimeter to near-infrared band) we are interested in, we always
have $\nu_B/\nu_\mathrm{obs}<10^{-3}$, and
$\nu_P/\nu_\mathrm{obs}<10^{-4}$. Therefore, although we consider
the Faraday conversion and rotation effects of magnetized plasmas,
we still assume photons of both natural modes propagating along
the null geodesics.

We use the ray-tracing code discussed in \citet{Huang07} to
determine the photon trajectory from a specific direction
as observed at infinity. Along the trajectory which crosses the
emission region, we record the four-wave-vector $k^\mu$, accretion
flow velocity $u^\mu$, electron temperature $T_e=\gamma_c m_e c^2
/k$, number density $n$, and three magnetic vector $\mathrm{B}^i$
at each line element. We convert the three magnetic vector
$\mathrm{B}^i$ into four-vector $b^\mu$ following  e.g.,
\citet{GMT03}. The four-vectors of the elliptical axes of
synchrotron emission $(e^{1\mu}, e^{2\mu})$ are calculated the
same way as that discussed by \citet{BB04}. The four-vectors of
the reference coordinates $(a^\mu, b^\mu)$ are calculated
according to the parallel transport in the general relativistic
theory with $(a^t,b^t)$ arbitrarily set as $(0,0)$ \citep{Chan83}.
The Stokes parameters are not conserved along the photon
trajectories due to gravitational effect.  However, the photon
occupation numbers  $\mathcal{N}_S = (\mathcal{N}_I,
\mathcal{N}_Q, \mathcal{N}_U, \mathcal{N}_V)$, defined as
$\mathcal{N}_S = \vec{S}/\nu^3$,  are Lorentz invariants.
Therefore, the radiative transfer equation (\ref{transferS}) can be
rewritten as
\begin{eqnarray}
\label{transferN}
  \frac{\rm d^{\quad}}{\rm d \ell^{\quad}} \mathcal{N}_S &=& \mathcal{E}
  \quad-\quad \bf {\mathcal{K}} \mathcal{N}_S,  \qquad (\mathcal{E} =
  \frac{\vec{\varepsilon}}{\nu^2},   \qquad \mathcal{K} = \nu
  \mathbf{K}),
\end{eqnarray}
with the differential affine parameter $\rm d \ell=\rm d
s/\nu_\mathrm{obs}$ and  the emission frequency $\nu=-u^\mu k_\mu$.

\section{SIMULATION RESULTS OF POLARIZATIONS}
\label{results}

In this section, we present our simulation results in detail.
The data we adopt are mainly from (multi-epoch) linear polarization
observations reported by \citet{Aitk00,Bowe05,Macq06,Marr06}, and
\citet{Ecka06}, marked with crosses in Panels (a), (b), and (c) of
Figure \ref{spec1}. Two
circular polarization data in Panel (d) are from \citet{Bowe01} and
\citet{Marr06}. The ties in X-ray band are from \citet{Baga01}.
Other data of luminosity shown in Panel (a) are the same adopted in
\citet{Huang08}.

\subsection{Configuration of the Magnetic Field}

We adopt the two-temperature magneto-rotational instability (MRI) driven
Keplerian accretion flow model of 
magnetized plasmas around the central black
hole in Sgr A* [see in \citet{Liu07, lfl07} for details].
A black hole mass of $M_\mathrm{BH}=4.1 \times 10^6 M_\odot$, where
$M_\odot$ is the solar mass, is recently derived from observations 
of the orbital motion of the short-period star S0-2 \citep{Ghez08}.
The corresponding model parameters include the ratio of the viscous
stress to the magnetic
field energy density $\beta_\nu$ [fixed at $0.7$, see Pessah et
al. (2006)], the ratio of the magnetic field energy density to the
gas pressure $\beta_p$, the electron heating rate indicated by a
dimensionless parameter
$C_1$, the mass accretion rate $\dot{M}$, the inclination angle $i$ and
the position angle  $\Theta$ of the axis perpendicular to the equatorial
plane of the accretion flow. In this model, thermal electrons are energised by
sound waves with the heating rate given by $\tau_{\rm ac}^{-1}=
c_S^2/3C_1H \langle v_e\rangle $, where $c_S$, $H$, and
$\langle v_e\rangle$ are the sound speed, the
scale height of the accretion flow, and the mean speed of the electrons,
respectively, and we
have assumed that the scattering mean free path of the electrons by the
sound waves is scaled with $H$. A comparison of this heating model with
other two-temperature models \citep{YQN03} can be found in \citet{lfl07}.

MHD simulations of the MRI show that the magnetic field is
dominated by its azimuthal component due to the shearing motion of
the Keplerian accretion flow. Previous pseudo-Newtonian modelling
of the sub-millimeter to NIR polarization of Sgr A* has assumed
that the magnetic field does not have vertical and radial
components \citep{MLC01,BML01,Liu07}. Here we consider more
realistic configurations.  Most MRI simulations adopt poloidal or
toroidal magnetic field loops confined to an initial torus of matter
\citep{HB02, GMT03}. Simulations with initial magnetic field lines
perpendicular to the disk plane tend to produce stronger magnetic fields
in the final quasi-steady state \citep{Pess06}. The field lines are dragged 
and twisted due to the Keplerian motion of the accretion flow. If there
is a large scale poloidal magnetic field, the field lines will
be nearly parallel (anti-parallel) to the velocity field in the upper half
to the equatorial plane and anti-parallel (parallel) in the lower half.  We
will adopt such a mean magnetic field configuration in the
following.

\subsection{The Fiducial Model}

\begin{figure}[ht]
\vspace{-0mm}
\begin{center}
\includegraphics[width=14.0cm]{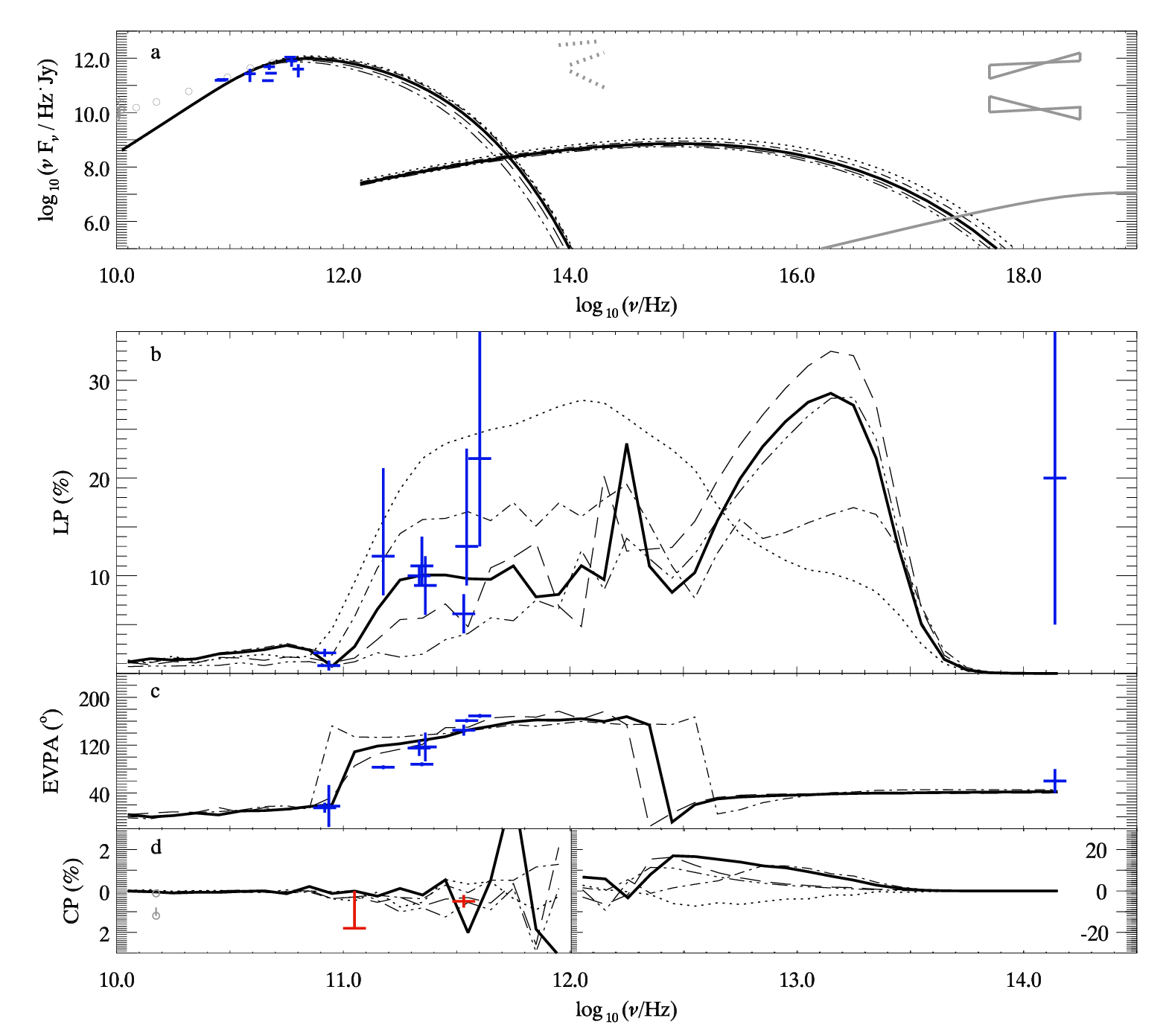}
\vspace{-5mm}\caption{ Simulated results from the fiducial model
with $\beta_\nu=0.7,\beta_p=0.4,C_1=0.47,\dot{M}=6\times 10^{17}
{\rm g \cdot s^{-1}}, i=40^\circ$, and $\Theta=115^\circ$ (solid
lines). Models with the same parameters except $i=20^\circ$
(3-dots-dashed lines), $30^\circ$ (long dashed lines), $i=50^\circ$
(dash-dotted lines), and $i=60^\circ$ (dotted lines) are shown for
comparison. Data are from
\citet{Aitk00,Bowe05,Macq06,Marr06,Ecka06}, \citet{Bowe01,Marr06},
\citet{Baga01},etc.. (a) spectrum of synchrotron, SSC radiation
(thick) and bremsstrahlung radiation (thin); (b) linear
polarization degrees; (c) EVPA; (d) circular polarization degrees.
\label{spec1}}
\end{center}
\end{figure}

We first obtain a fiducial model to the spectrum, linear
polarization (LP) fraction, electric vector position angle (EVPA),
and circular polarization (CP) fraction with $\beta_p=0.4$,
$C_1=0.47$, $\dot{M}=6 \times 10^{17} \mathrm{g} \cdot
\mathrm{s}^{-1}$, $i=40^\circ$, and $\Theta=115^\circ$. No
external Faraday rotation measure is introduced. The corresponding
results are shown by the solid lines in Panels (a),(b),(c), and
(d) of Figure \ref{spec1}, respectively. It is consistent with the
fiducial model in Huang et al. (2008). This Keplerian accretion flow
reproduces the emission in the millimeter to sub-millimeter bump
of Sgr A*, but underestimates the emission in the NIR and X-ray
bands. The thick solid lines in Figure \ref{spec1} (a) represent
the synchrotron (to the left) and synchrotron self-Comptonization
(SSC, in the middle) radiation components, and the thin solid line
to the right represents the bremsstrahlung radiation. This result
is different from that of Liu et al. (2007), where they are able
to reproduce the observed NIR flux level. In their model, an
un-polarized jet/outflow component were introduced, which
effectively suppresses the LP below 100 GHz \citep{Liu07}. We
don't have such a component here. The onset frequency of the LP
therefore constrains the model parameter space so that the model
predicted NIR and X-ray fluxes are more than one order of
magnitude lower than the observed values. The quiescent-state X-ray
emission is extended and has been attributed to thermal emission from
plasmas near the capture radius \citep{Baga03, Q04, Xu06}. The X-ray
emission from our accretion flow at small radii should be lower than the
observed value. However, there is evidence for a quiescent-state NIR
emission from the direction of Sgr A* \citep{D09}. This emission can
be caused by smaller flares or a quasi-stationary emission component. If
it is indeed powered by the accretion flow at small radii, it must be
dominated by emission regions within $6GM_{\rm BH}/c^2$, where $G$ is
the gravitational constant and $c$ is the speed of light, or associated
with outflows, which may also be responsible for the long wavelength radio
emissions.    \footnote{The black hole does not have a spin in
	   our pseudo-Newtonian model. The accretion disk has an inner
	   boundary radius of $6GM_{\rm BH}/c^2$.}

As shown in Figure \ref{spec1} (b) and (c), the LP fraction is a
few percent in the centimeter bands. It can be reduced to the
observed level either by depolarization of an external medium or
through the dominance of an un-polarized emission component from a
jet/outflow component. In this band, only the red-shifted side of
the accretion flow is optically thin. Hence, the EVPA is
perpendicular to the accretion flow axis projection, since the
magnetic field is toroidal and the emission is dominated by the
extraordinary mode. When the observational frequency increases to
$\sim$100GHz, the LP degree increases because the Faraday rotation
and Faraday conversion coefficients decrease with the observation
frequency $\nu$ as $\nu^{-2}$ and $\nu^{-3}$, respectively. At the
same frequency, the near and far sides of the accretion flow become
optically thin and 
dominate the polarized emission so that EVPA preforms a $\sim 90^\circ$
flip to be parallel to the disk axis projection. In the 
sub-millimeter band, the LP degree increases to $\sim 10 \%$,
explaining observations well. Furthermore, the EVPA has a small
increase of $\sim 20^\circ$ in this band. With further increasing
of the observational frequency, the blue-shifted side finally
become optically thin at $\sim2-3$ THz, making the whole disk
optically thin with its blue-shifted side being the dominant emission
region. Thus, the EVPA flips back to be perpendicular to the axis
projection again. In the quiescent state discussed in this paper,
there is almost zero polarization in the NIR band because the
non-polarized SSC emission component dominates. However, the EVPA
is still meaningful for flaring activities dominated by the
synchrotron emission.

The predicted CP fraction is shown in Figure \ref{spec1} (d).
The CP degree is nearly zero in the centimeter bands, so that the observed
CP fraction of several percents (Bower et al. 2002) might be attributed
to a jet/outflow component (Beckert \& Falcke 2002).  At sub-millimeter
and shorter 
wavelengths, however, the CP amplitude increases gradually, then
significantly to greater than $10 \%$ at $\sim$1 THz.
We show the CP fraction in different scales for clarity of presentation here.
Notice that the LP fraction also has some oscillations in the same
band. These high LP and CP may be due to the combination of the general
relativity and birefringence effects, and can be tested with
future THz polarization observations.

Interestingly in the millimeter to sub-millimeter band, 
the CP degree is mostly negative, i.e., the polarization is left-handed. 
Here, we adopt two observational data for CPs, a limit of $-1.8\%$ at 112 GHz
reported by \citet{Bowe01} and a detection of $\sim -0.5\% \pm
0.3\%$ at 340GHz by \citet{Marr06} (both marked with red crosses).
Although they mentioned that the detections were quite uncertain
due to weather conditions and instrumental effects, the
left-handed properties seemed to be real. Our model therefore
reproduces the CP observations. However, we notice that the CP
degree is very sensitive to the magnetic field structure and model
parameters. Thus, our predictions here should be interpreted with
precautions.

\subsection{The Inclination Angle Dependence}

The dependence of these results on the inclination angle $i$ is
also shown in Figure \ref{spec1} with the corresponding models
indicated in the legend of Panel (b). The other model parameters
have been fixed here. Results for $i=20^\circ$, $30^\circ$,
$50^\circ$, and $60^\circ$ are shown with the 3-dots-dashed, long
dashed, dot-dashed, and dotted lines, respectively. Different from
the pseudo-Newtonian cases [see \citet{Liu07}], the changes in the
inclination angle slightly affect the centimeter polarization and
high frequency flux density. This is due to two effects:
absorption of emission from the edge of accretion flow and
amplification of emission by the GR light bending.

However, the inclination angle affects the LP degree
significantly. Generally speaking, the onset frequency of high LP
decreases and the LP degree in sub-millimeter band increases with
the increase of the inclination angle. This is because the
projected magnetic field lines are close to parallel lines in the
nearly edge-on case while becoming circular helices in the nearly
face-on case. Parallel magnetic field lines result in the highest LP
degree (up to $30 \%$ in sub-millimeter band), while circular
helices result in a zero LP degree due to the rotation symmetry of
the flow. The observed LP degrees in the sub-millimeter band
remain stable at $\sim 10\%$ since its discovery \citep{Aitk00},
which provides a constraint of $\sim 40^\circ \pm 10^\circ$ on the
inclination angle of the Keplerian accretion flow.

For the EVPA, it flips at a lower frequency for the first time and
a higher frequency again. This is because the near side of the inclined
disk dominates the emission in a wider band. For the
$i=60^\circ$ case, the near side dominates after the first flip,
and the second flip doesn't appear until the frequency increases to 
the NIR band. The second flip may disappear in the band we are interested in 
for higher values of the inclination angle. 
The second flip between sub-millimeter band to NIR band is
quite important. A stable difference of $\sim 80^\circ$ in the
EVPAs between 230GHz and $\lambda$ 2.2$\mu$m has been observed,
which suggests a stable geometry for the accretion plasma around
Sgr A* if the NIR emission also originates from the disk.
Comparing the fiducial model and the other four cases shown in
Fig.\ref{spec1}, a mildly-inclined $i\sim 40^\circ$ accretion flow
appears to better reproduce such a difference in the EVPA caused
by changes in optical depth across the emission structure.
However, \citet{Meye07} derived a highly-inclined $i \gtrsim
70^\circ$ disk to obtain significantly variable light-curves in
the NIR band. The NIR flare emission was associated with a bright
spot orbiting the black hole at the last stable orbit of a disk
with a toroidal magnetic field similar to our magnetic field
configuration. They also assumed that the sub-millimeter emission
comes from a jet to explain the EVPA difference, while the
jet/outflow component in our model only dominates the emission in
millimeter and longer wavelengths. Therefore, the difference of
inclination angle predictions is caused by different assumptions
of the geometry for the emitting region, which can be checked by
future VLBI observations in the sub-millimeter band.

\subsection{The Position Angle Dependence}

The observed EVPA is related to the position angle ($\Theta$) of
the accretion flow axis. Observationally, the EVPAs in the
sub-millimeter band show frequent variability, presumably
associated with flares. At 230 GHz, however, it was detected as
$\sim 117^\circ \pm 24^\circ$, showing relatively stable values,
during several epochs within more than one year \citep{Bowe05}.
\citet{Marr07} also reported comparable values of the EVPA at
230GHz. At $\lambda$ 2.2$\mu$m, on the other hand, the EVPA is
reported to be stable with a value of $60^\circ \pm 20^\circ$
\citep{Meye07}. In the fiducial model, we find $\Theta=114^\circ$
so that the predicted EVPA can give a best fit to both
measurements at 230GHz and $\lambda$ 2.2 $\mu$m. In practice,
values in the range of $\sim 115^\circ \pm 20^\circ$ are all
acceptable. Interestingly, we find that the predicted first flip
in the EVPA also fits the multi-epoch detections at 86 GHz well
\citep{Macq06}. Moreover, \citet{Marr07} observed the EVPA at
340GHz simultaneously with that at 230GHz and reported an averaged
$\sim 30^\circ$ increase from 230 GHz to 340 GHz.  Our model also
reproduces this increase in the sub-millimeter band.
\citet{Meye07} derived a $\Theta$ of $\sim 110^\circ$, which is
consistent with ours, although their emission model is quite
different.

\begin{figure}[ht]
\vspace{-0mm}
\begin{center}
\includegraphics[width=14.0cm]{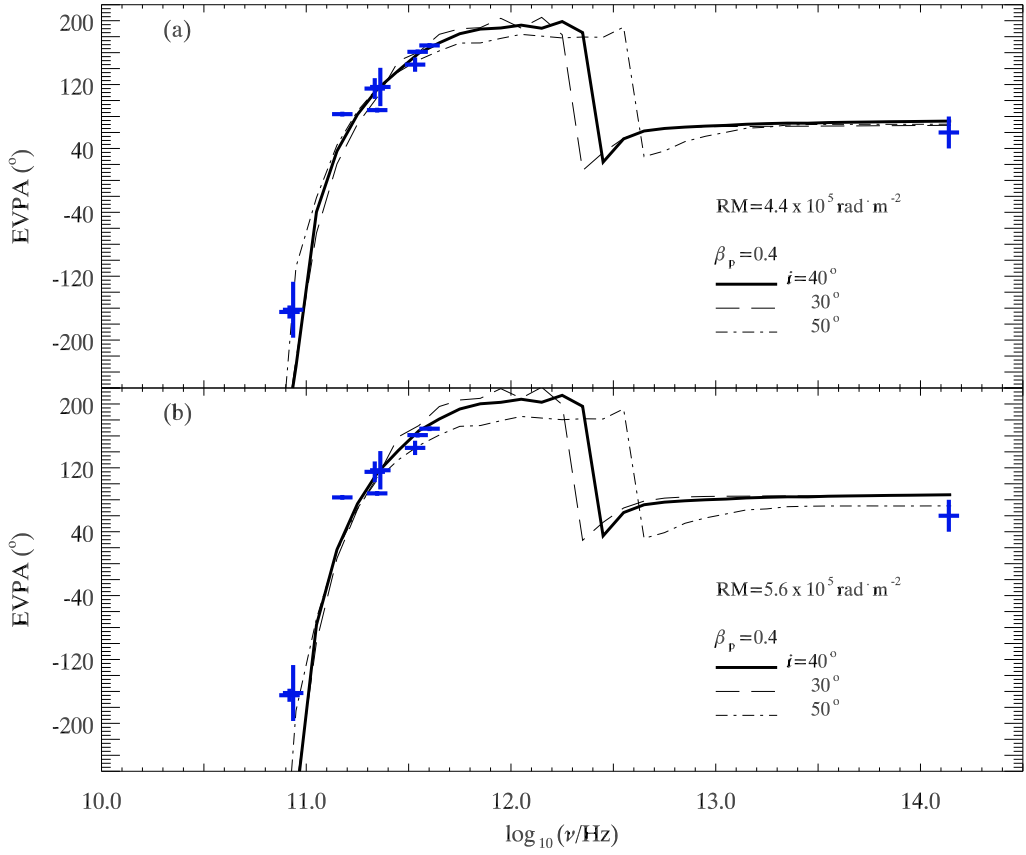}
\vspace{-5mm}\caption{ Results of EVPA three models in
Fig.\ref{spec1} with external rotation measure of (a) $4.4 \times
10^5 {\rm rad} \cdot {\rm m}^{-2}$ \citep{Macq06} and (b) $5.6
\times 10^5 {\rm rad} \cdot {\rm m}^{-2}$ \citep{Marr07}
considered. \label{RM}}
\end{center}
\end{figure}

Notice that we have assumed that all the observed EVPAs are
intrinsic to the accretion flow, i.e., the depolarization and Faraday
rotation only occur in the accretion region of Sgr A*.  It is likely
that the emission from the accretion flow may experience depolarization
and Faraday rotation by
an external medium described with a rotation measure (RM) $=4.4 \times
10^5 \mathrm{rad} \cdot \mathrm{m}^2$ derived from 
millimeter to sub-millimeter observations \citep{Macq06}. In Panel
(a) of Figure \ref{RM}, we plot the corresponding EVPAs of three
well-fit models in Figure \ref{spec1}, with different position
angles. Multi-epoch observations of the EVPA from $\lambda=$ 3 mm
to $\lambda=$ 2.2 $\mu$m can be reproduced with $\Theta=147^\circ$
for the fiducial model. Fits with another RM of $5.6 \times 10^5
\mathrm{rad} \cdot \mathrm{m}^2$ derived  by \citet{Marr07} from
their simultaneous observations at $230$ GHz and $350$ GHz are
plotted in Panel (b). $\Theta=159^\circ$ for the fiducial model.
Obviously, the position angle $\Theta$ depends on the external
rotation measure. More observations in the sub-millimeter and NIR
bands will show whether an external rotation measure is necessary.

\subsection{Dependence on the Mass Accretion Rate}

\begin{figure}[ht]
\vspace{-0mm}
\begin{center}
\includegraphics[width=14.0cm]{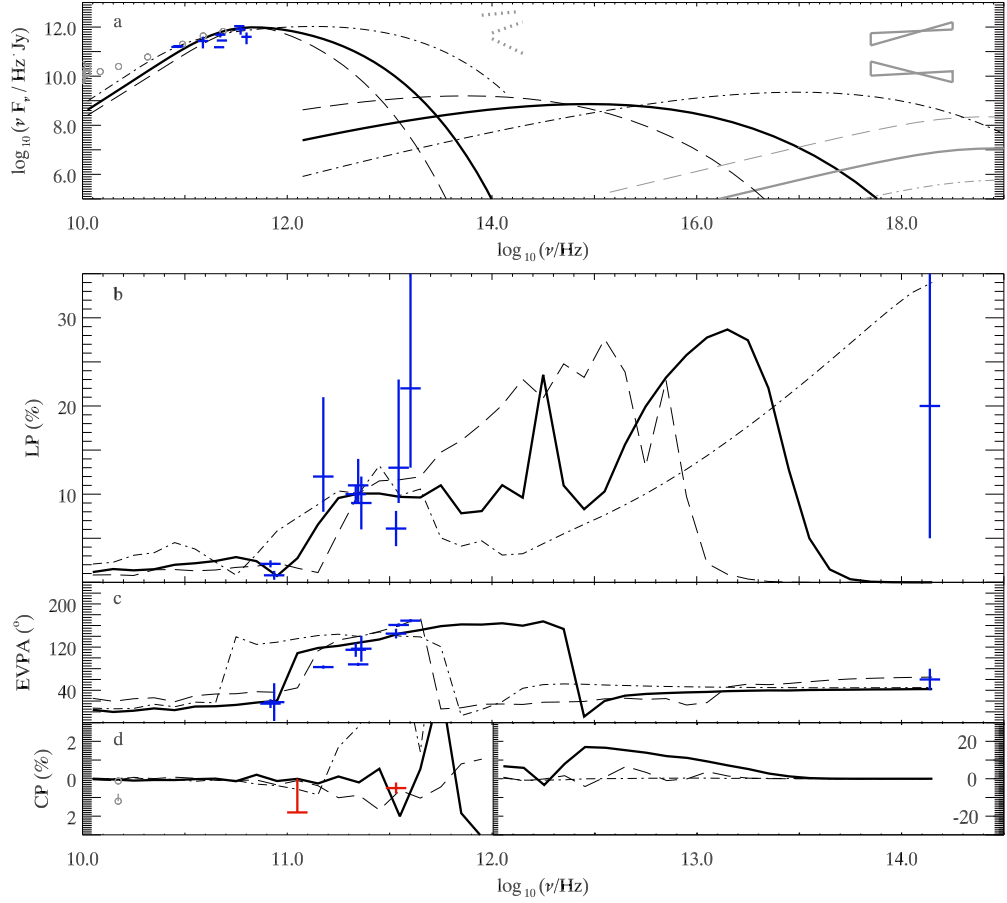}
\vspace{-5mm}\caption{ Simulated results for models with
$\beta_p=0.2$, $\dot{M}=1.5\times 10^{18}$g s$^{-1}$ (long-dashed),  $\beta_p=0.4$ (solid), and
$\beta_p=0.7$,  $\dot{M}=2\times 10^{17}$g s$^{-1}$ (dash-dotted), where
 $\beta_p \cdot C_1 = 0.4 \cdot 
0.47$, and the luminosity in sub-millimeter band does not
change significantly. \label{spec2}}
\end{center}
\end{figure}

For these radiatively inefficient accretion flows, the gas
pressure is dominated by proton and ions, which have a relatively
high temperature Liu et al. (2007). Pessah et al. (2006) showed
$\beta_\nu\simeq 0.7$. Then both the viscosity (and therefore the
mass accretion rate) and the magnetic field energy density are
proportional to $\beta_p$. The mass accretion rate therefore
determines the amplitude of the magnetic field. For a given mass
accretion rate, the density is inversely proportional to
$\beta_p$. Figure \ref{spec2} shows the dependence of the results
on the mass accretion rate, where $\beta_p$ and $C_1$ are also
adjusted to reproduce the millimeter and sub-millimeter spectrum.
For $\dot{M}=1.5 \times 10^{18} \mathrm{g} \cdot \mathrm{s}^{-1}$,
$\beta_p=0.2$, $C_1=0.94$. The density profile increases by a
factor of 5 with respect to the fiducial model. And for a lower
accretion rate of $\dot{M}=2 \times 10^{17} \mathrm{g} \cdot
\mathrm{s}^{-1}$, $\beta_p=0.7$, $C_1=0.269$. The density profile
decreases by a factor of 5.3 with respect to the fiducial model.
The inclination angle of the accretion flow is fixed at $40^\circ$
and $\Theta$ is obtained by fitting observations at $1.3$ mm and $2.2 \mu$ m
wavelengths. It can be seen that with the increase of $\dot{M}$
and therefore the magnetic field $B$ and $n$, a lower electron
temperature is needed to reproduce the spectrum. This leads to a
lower cutoff frequency for both the synchrotron and SSC
components. The emission is also strongly self-absorbed so that
the onset frequency of strong LP occurs at about 150 GHz. Opposite
effects can be seen with the decrease of $\dot{M}$ and there is a
higher X-ray emission flux due to the SSC. However, the X-ray flux
is still one order of magnitude lower than the observed
quiescent-state thermal X-ray emission from a large scale
accretion flow (Xu et al. 2005). Even higher NIR and X-ray
emission may be produced with further decrease of the mass
accretion rate and increase of the electron temperature. However,
radio emission will show strong LP. An un-polarized emission
component needs to be introduced to suppress the LP to the
observed level (Liu et al. 2007).

Moreover, for lower values of $\dot{M}$, the high amplitude of the CP
degree starts at lower frequencies, similar to the high LP
degrees, e.g., the dash-dotted line shown in Figure \ref{spec2}
(d) with $\beta_p=0.7$. Current observations of the LP and CP
therefore constrain the $\dot{M}$ to be near $6\times 10^{17}$ g
s$^{-1}$ with an uncertainty of a factor of $\sim 2$. \footnote{This
result is for the pseudo-Newtonian potential without a black
hole spin. The range of $\dot{M}$ will be different in a Kerr metric
with a significant spin.}

\subsection{Degeneracy Between the Electron Heating and Magnetic Field}

\begin{figure}[ht]
\vspace{-0mm}
\begin{center}
\includegraphics[width=14.0cm]{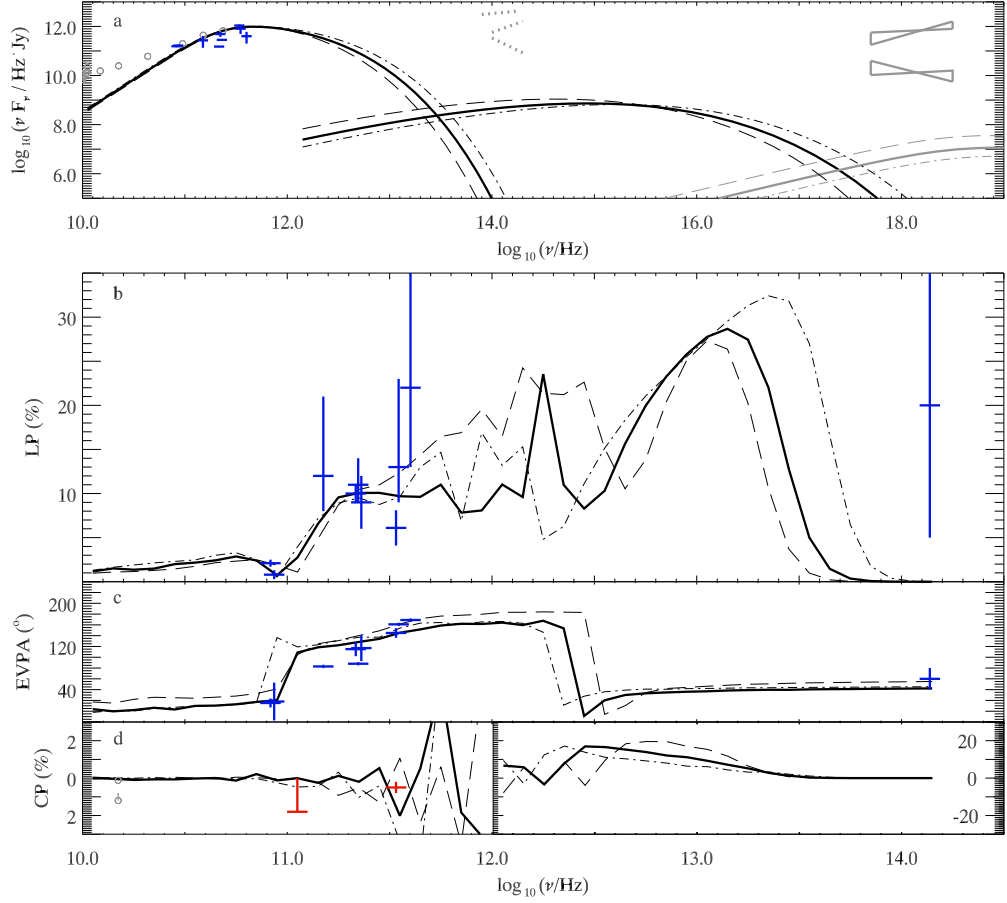}
\vspace{-5mm}\caption{Simulated results from models with
$\beta_p=0.2$, $C_1=0.8$ (dashed) and $\beta_p=0.7$
$C_1=0.33$ (dot-dashed) with $\dot{M}=6\times
10^{17} {\rm g \cdot s^{-1}}$ and $i=40^\circ$ unchanged.
\label{betapC1}}
\end{center}
\end{figure}

We have shown that the emission spectrum and polarization give
good constraints on the orientation of the accretion flow and the
mass accretion rate. However, the constraint on the electron
heating rate and the magnetic field parameter $\beta_p$ is
generally poor even with $\beta_\nu$ fixed. Figure \ref{betapC1}
shows several reproductions to the observed spectrum and
polarization with $\dot{M} = 6\times 10^{17}$ g s$^{-1}$,
$\beta_p=0.2$; $C_1=0.8$ (long-dashed lines), and $\beta_p=0.7$;
$C_1 = 0.33$ (dashed-dotted lines). The current observations will
not be able to distinguish these models. However, there are
changes in the SSC and bremsstrahlung components. These can be
understood as the following.

For a given $\dot{M}$, the magnetic field is fixed because both of
them are proportional to $\beta_p N_0 T_p$, where $T_p$ is the
proton temperature and does not change with $\beta_p$.  With the
increase $\beta_p$, we increase the viscosity and therefore the
radial velocity. The density will decrease as $N_0 \propto
1/\beta_p$. For a given optically thin synchrotron luminosity
required to reproduce the millimeter and sub-millimeter spectral
hump, $N_0 B^2 T_e^2$ should not change.  Therefore the electron
temperature $T_e$ should be proportional to $\beta_p^{1/2}$. Then
with the increase of $\beta_p$, the synchrotron spectrum should be
slightly harder, i.e., its spectral cutoff $\propto B T_e^2
\propto \beta_p$ should shift toward higher frequencies. We
therefore expect less low frequency SSC flux density due to the
decrease of the electron density. However the SSC component should
cut off at a higher frequency, which is proportional to $T_e^2
\propto \beta_p$. The SSC luminosity, which is proportional to
$N_0 T_e^2$, however does not change. To make the SSC cut off in
the Chandra X-ray band, we will need to increase $\beta_p$ by a
factor of $\sim 10$. However, this will reduce the flux density by
the same factor. Therefore the SSC component cannot have
significant contribution to the observed quiescent state X-ray
emission. The bremsstrahlung flux density scales with
$N_0^2/T_e^{1/2}\propto \beta_p^{-9/4}$. A very low value of
$\beta_p$ is required for this component to have significant
contribution in the X-ray band. There is therefore a degeneracy
between the electron heating rate and the magnetic parameter. By
introducing a spin parameter to the black hole, the emission
volume will be reduced and therefore enhancing the SSC X-ray flux.
One may be able to break this degeneracy with the quiescent state
NIR and X-ray flux densities.

\subsection{Polarized Images in the Millimeter and Sub-Millimeter Bands}
\label{image}

\begin{figure}[ht]
\vspace{-0mm}
\begin{center}
$\lambda = 3.5$mm \\
\includegraphics[width=14.0cm]{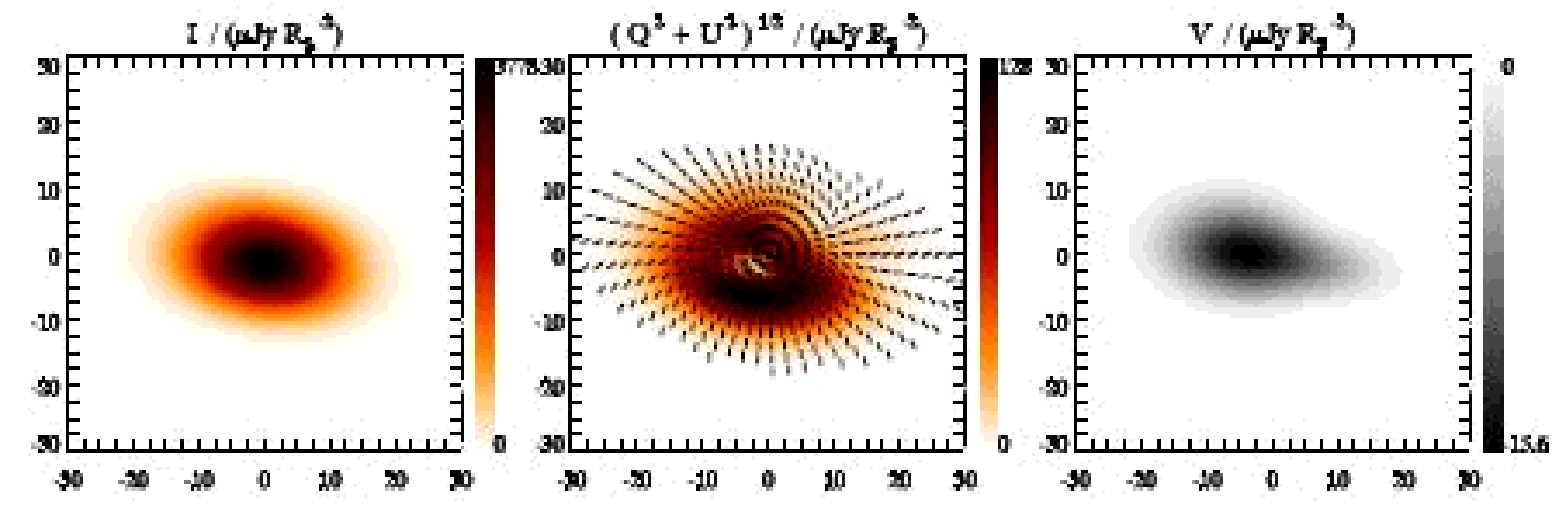} \\
$\lambda = 1.3$mm \\
\includegraphics[width=14.0cm]{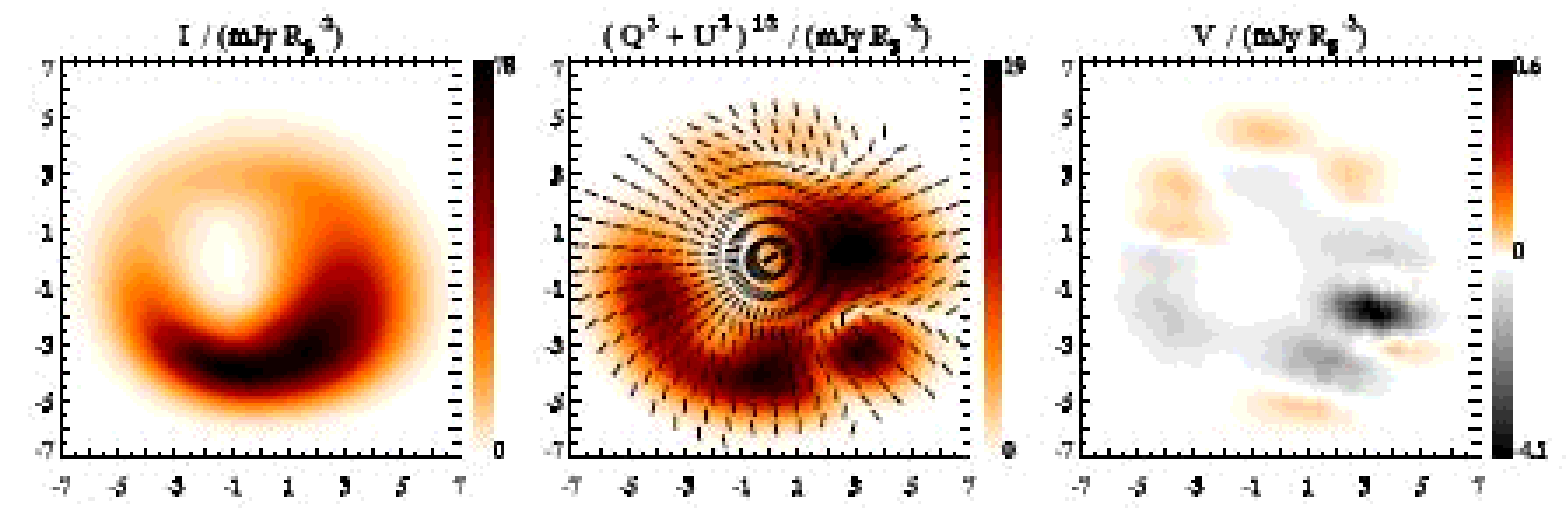} \\
$\lambda = 0.86$mm \\
\includegraphics[width=14.0cm]{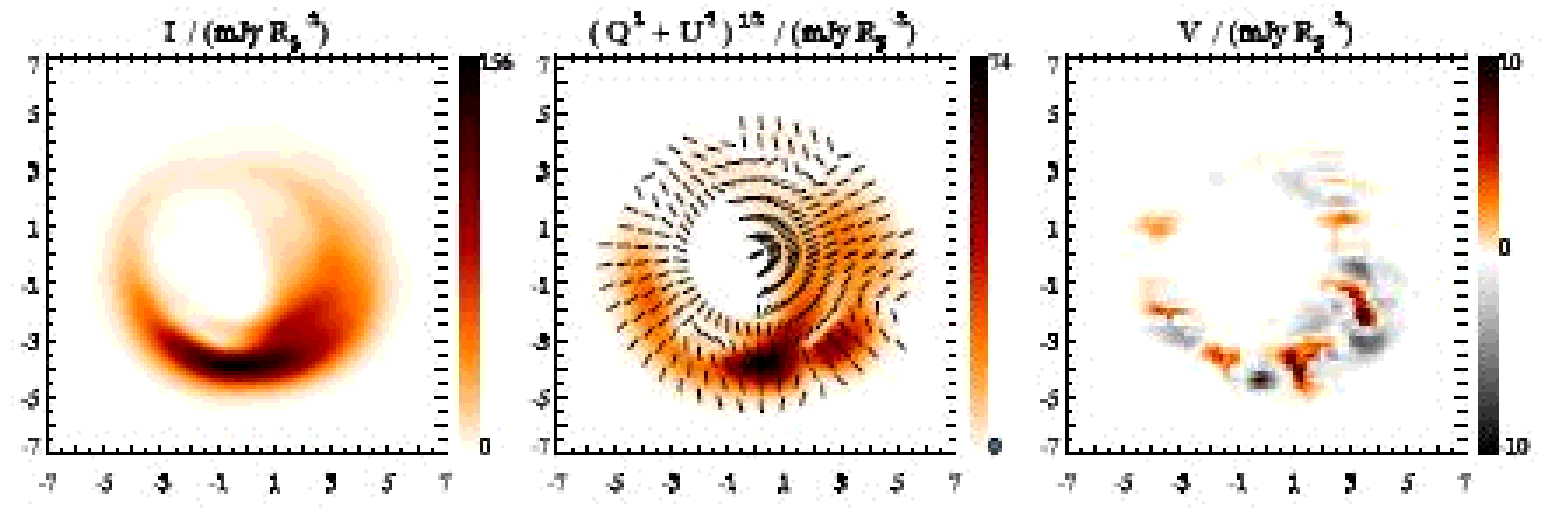} \\
$\lambda = 0.6$mm \\
\includegraphics[width=14.0cm]{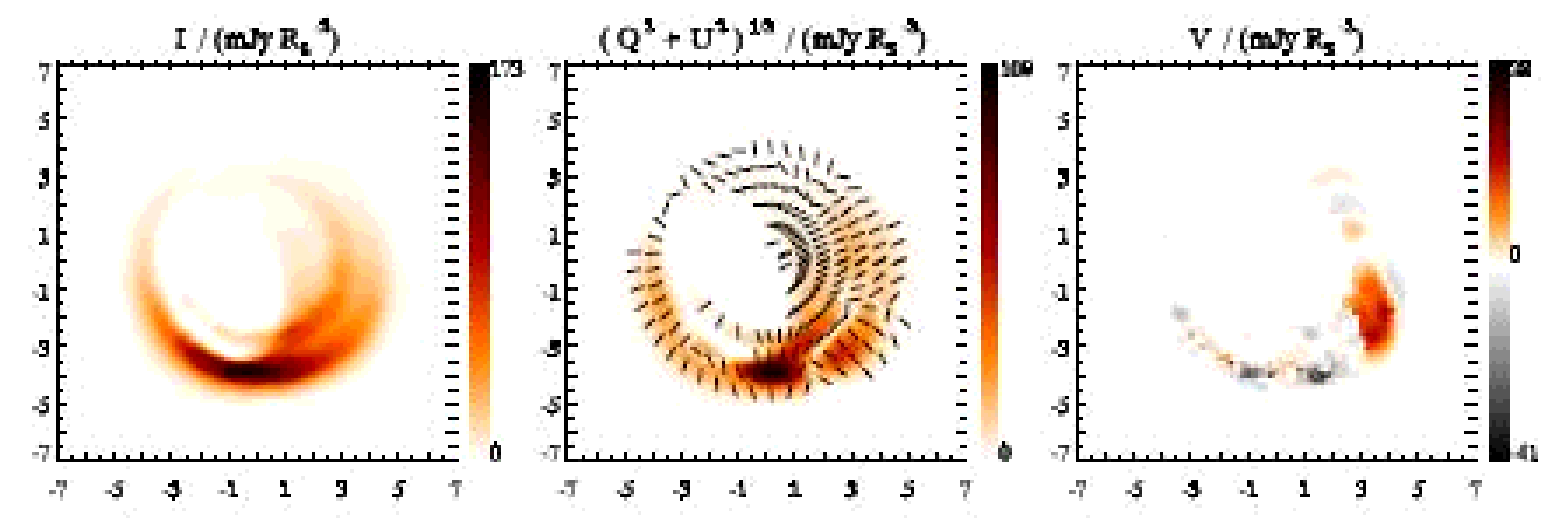} \\
\vspace{-5mm}\caption{
Images of total intensity ($I$), LP emission ($\sqrt{Q^2 + U^2}$), and CP emission ($V$)
predicted for Sgr A* at observational wavelengths of 3.5 mm, 1.3 mm,
 0.86 mm, and 0.6 mm, from top to bottom, respectively.
In images of LP emission, black spurs represent the averaged EVPAs in
 the nearby regions. 
In images of CP emission, the left-handed and right-handed regions are
 respectively presented in grey and red. 
\label{SD}}
\end{center}
\end{figure}

Sgr A* is embedded in the inter-stellar medium (ISM) on the
Galactic plane. Emission from the accretion flow will experience
significant scattering, which smoothes the image without changing
its total flux density and polarization properties. We adopt an
elliptical Gaussian structure for the scattering screen with a
full width of half maximum (FWHM) along its major and minor axis
being $\vartheta_\mathrm{maj}=(1.39 \pm 0.02) (\lambda/ 1 {\rm\ 
cm})^2$ mas and $\vartheta_\mathrm{min}=(0.69 \pm 0.06) (\lambda/
1 {\rm\  cm})^2$ mas, respectively, and a position angle $\sim
80^\circ$ \citep{Shen05}. The size of the scattering screen
becomes smaller with the decrease of the observation wavelength.
In practice, the black hole shadow structure is significantly
washed out in the millimeter and longer wavelength band by this
scattering. However, the scattering becomes less important in
sub-millimeter band. We plot images of the total emission $I$, LP
:$\sqrt{Q^2 + U^2}$, and CP: $V$ of the fiducial model at several
wavelengths $\lambda=$ 3.5mm, 1.3mm, 0.86mm, and 0.6mm in Figure
\ref{SD}. The black dashes in the LP emission represent the EVPAs
$\bar{\chi}$. The left-handed and right-handed regions in the CP
emission are respectively represented with grey and red color.

The image of the total emission at $\lambda=$ 3.5 mm is smeared by the
screening and has 
an $\sim 80^\circ$-oriented elliptical Gaussian structure
with the major FWHM of $\sim (0.189 \pm 0.003)$mas. This size is
consistent with but somewhat smaller than the observation of $\sim 0.21$
mas reported in \citet{Shen05}. The LP and CP emission at $\lambda$ 
3.5 mm are also smoothed significantly. The spurs mark the EVPAs, mostly in
the radial direction as a result of the toroidal magnetic field
configuration. The CP emission is dominated by a grey region
representing a left-handed polarization.

Clear black hole shadow structures can be found in images of the total
emission at 1.3 mm or shorter wavelengths.
Interestingly, clumpy patterns are seen in the sub-millimeter LP with
bright and faint regions, and in the CP emission with their
left-handed and right-handed regions. Such complex patterns are
results of the gravitational light bending and the
birefringence effects posing new challenges for future
polarization-sensitive VLBI observations.

\section{SUMMARY AND DISCUSSION}
\label{discuss}

In this paper we explore the parameter space of the
two-temperature MRI driven Keplerian accretion flow model in a
non-spin pseudo-Newtonian potential for the time averaged
millimeter and shorter wavelength emission from Sgr A* (Liu et al.
2007). The model reproduces the observed emission spectrum and
polarization with an inclination angle of $40\pm 10^\circ$ and a
mass accretion rate of $\sim 6\times 10^{17}$ g s$^{-1}$. The
former is mostly determined by the amplitude of the LP fraction
and the latter is well-constrained by the onset frequency of
prominent LP. The LP is low in the centimeter wavelength due to
strong self-absorption in the emission region. The orientation of
the accretion flow projected on the sky depends on the  amplitude
of an external Faraday rotation measure. And the model predicted
NIR and X-ray fluxes are more than one order of magnitude lower
than the observed low flux levels. Although nearly identical
millimeter and sub-millimeter spectrum and polarization can be
obtained by adjusting the electron heating rate and the magnetic
parameter, none of these models can produce NIR and X-ray fluxes
comparable to the observed values. This strongly suggests that the
black hole is rotating so that the last stable orbit can be
smaller, resulting in more SSC and bremsstrahlung emission. We are
in the process of this study and the results will be reported in a
separate paper.

The time averaged source size measurements provide critical
constraints on the model. Although we showed that our model are
consistent with current observations by simulating the images of
the accretion flow at different wavelengths. Given the challenges
in imaging the black hole with the global VLBA, a direct fit to
the observed visibility is needed to fully utilize the observed
information \citep{Doel08}. 

Very rich information is contained in the flare observations.  To
fully explore the implication of these observations, one needs to
carry out GRMHD simulations of the accretion flow. Our currently
study of the time averaged properties will lead to good
constraints on the properties of the plasmas near the black hole.
These will be helpful for setting up the simulations to recover
the essential observations. Besides GRMHD simulations, the
kinetics of electron heating and acceleration must also be
addressed (Liu et al. 2004). The GR ray-tracing treatment of
polarized radiation transfer we have here in this paper, GRMHD
simulations, and a kinetic model for the electron acceleration
should be combined to develop a self-consistent model for the
emission structure near the black hole. In light of continuous
high resolution and sensitivity observations, these theoretical
developments will be essential to uncover the nature of the black
hole and its interaction with the plasmas feeding it.

\acknowledgments

This work was supported in part by the National Natural Science
Foundation of China 
(grants 10573029, 10625314, 10633010, 10821302, 10733010, 10673010, and
10573016)  
and the National Key Basic Research Development Program of China
(No.2007CB815405 and 2009CB824800). 
SL is supported by a Marie Curie Fellowship under the EC's SOLAIRE
Network at the University of Glasgow (MTRN-CT-2006-035484) and an IGPP grant from LANL.
ZQS is supported by the Knowledge Innovation Program
of the Chinese Academy of Sciences (Grant No. KJCX2-YW-T03) 
and the Program of Shanghai Subject Chief Scientist (06XD14024). 
YFY is supported by Program for New Century
Excellent Talents in University. 
LH thanks G.-X. Li for the assit in programming.


\vspace{5cm}

\end{document}